\documentclass{article}
\usepackage{graphics,wrapfig,epsfig}

\begin{document}

\begin{center}

{\bf
\title "The equilibrium state of the dense electron-nuclear plasma in the self-gravitational field.
The stellar mass distribution and stellar magnetic fields.}

\bigskip

\bigskip
\author "B.~V.~Vasiliev
\bigskip

Institute in Physical-Technical Problems, 141980, Dubna, Russia
\bigskip

{vasiliev@dubna.ru}
\end{center}

\bigskip

\begin{abstract}
The equilibrium of dense plasma in a self-gravitation is
considered. The obtained results radically distinguish from the
point of view which is commonly accepted in the astrophysical
society. It is important that all these results were obtained
without any disputable speculative assumptions. They were obtained
on the standard physical base by standard formal methods. The
novelty of the obtained results is based on a rejection of the
oversimplified ideal gas approximation which is usually accepted
for a star interior description and on a taking into consideration
the electron-nuclear plasma features. It was shown that there is
the minimum for plasma energy at a density and a temperature which
determines the equilibrium state of plasma in the self-gravitation
at zero gradient of the general parameters of plasma. This effect
plays an important role for astrophysics. It enables to explain
the mechanism of the star magnetic field generation and to make a
prediction for the spectrum of star masses with a quite
satisfactory agreement for the observation data.
\end{abstract}

\bigskip

\bigskip
\bigskip

\section{The current hypothesis}

    Now it is conventionally accepted to think that the density
and the temperature of a star interior substance are growing
depthward in a star and can amount to tremendous values at its
central core. It seems that this growing is a necessary condition
of an equilibrium of a star substance in the self-gravity.

    The substance exists as hot dense electron-nuclear plasma at
high pressures and temperatures of a star interior. At this
condition in the zero approximation, plasma can be considered as a
Boltzmann ideal gas with energy
\begin{equation}
E=\frac{3}{2}kT N \label{eB}
\end{equation}
where $N$ is the particle number.

Since the direct inter-nuclear interaction in plasma is small, it
can be neglected and one can write the equilibrium equation in the
form \cite{1}:

\begin{equation}
\mu_e+m'\psi=const\label{equ1}
\end{equation}
where $\mu_e$ is the electron chemical potential,
$m'={\gamma}/{n_e}$ is the mass of the substance related to one
electron, $n_e$ is the electron gas density, $\gamma$ is the mass
density of plasma, $\psi$ is the Newton gravitational potential.
Because $\Delta \psi=-4\pi G \gamma$ in a spherically symmetric
case, Eq.({\ref{equ1}}) is reduced to

\begin{equation}
\frac{1}{r^2} \frac{d}{dr}\biggl(r^2 \frac{d\mu_e}{dr}\biggr)
=-4G\pi\gamma m'.\label{equ2}
\end{equation}
The electron gas chemical potential
\begin{equation}
\mu_e= kT ln \biggl[\frac{n_e}{2}\biggl(\frac{2\pi \hbar^2}{m
kT}\biggr)^{3/2}\biggr]\label{muB}
\end{equation}
is the function of the particle density and the temperature only
\cite{1}. At the consideration of the hot non-degenerate plasma in
an ideal gas approximation, one can deduce an unambiguous
conclusion from the equilibrium equation (Eq.({\ref{equ2}})): the
balance of plasma particles in the self-gravitation demands
temperature and density increasing depthward in a star.

The interparticle interaction is completely neglected in the ideal
gas approximation by definition. It is an oversimplified
assumption for electron-nuclear plasma, although at the first
sight it is acceptable because the interparticle interactions are
small in comparison with the ideal gas energy (Eq.({\ref{eB}})).
But the allowance of these interactions has a principal importance
because they form a stable equilibrium state of hot plasma. It
falls out of the consideration when the ideal gas approximation is
used.

\section{The density and temperature of a hot dense plasmas in the equilibrium state}

\subsection{The steady-state density of a hot dense plasma}

There are two main characteristic features which we must take into
account at hot dense plasma consideration. The first of them is
related to the quantum  properties of the electron gas. The second
feature is concerned with the presence of positively charged
nuclei inside the electron gas of plasma.

\subsubsection{The correction for Fermi-statistics}

The estimation of the electron gas energy in the Boltzmann case
$(kT\gg E_F)$ can be obtained by a series expansion of the full
energy of the non-relativistic Fermi-particle system \cite{1}:

\begin{equation}
E=\frac{2^{1/2}V m_e^{3/2}}{\pi^2 \hbar^3} \int_0^\infty
\frac{\varepsilon^{3/2}d\varepsilon}
{e^{(\varepsilon-\mu_e)/kT}+1}, \label{eg}
\end{equation}
where $\varepsilon$ is the energy of a particle. In the Boltzmann
case, $\mu_e<0$ and $|\mu_e/kT|\gg 1$ and the integrand at
$e^{\mu_e/kT}\ll 1$ can be expanded into a series according to
their powers $e^{\mu_e/kT-\varepsilon/kT}$.

As a result, taking into account its quantum properties, the hot
electron gas full energy obtains the form \cite{1}

\begin{equation}
E=\frac{3}{2} kT
\biggl[1+\frac{\pi^{3/2}}{4}\biggl(\frac{a_0e^2}{kT}\biggl)^{3/2}n_e\biggr],
\label{cef}
\end{equation}
where $a_0=\frac{\hbar^2}{me^2}$ is Bohr radius.

It is important to underline that the correction for
Fermi-statistics is positive, because it takes into account that
an electron cannot occupy energetic positions filled by other
electrons and the resulting pressure is more than the pressure of
an ideal gas at exactly the same density and temperature.

\subsubsection {The correction for a correlation of charged
particles in plasma}

At high temperature, the plasma particles tend to uniform space
distribution. At this limit, the energy of ion-electron
interaction tends to zero. Some correlation in space distribution
of particles arises as the positively charged particle groups
around itself preferably particles with negative charges and vice
versa. It is accepted to estimate the energy of this correlation
by the method developed by Debye-H$\ddot u$kkel for strong
electrolytes \cite{1}. The energy of a charged particle inside
plasma is equal to $e\varphi$, where $e$ is the charge of a
particle, and $\varphi$ is the electric potential induced by other
particles  on the particle under consideration.

This potential inside plasma is determined by the Debye law
\cite{1}:

\begin{equation}
\varphi(r)=\frac{e}{r} e^{-\frac{r}{r_D}}\label{vr}
\end{equation}
where the Debye radius is

\begin{equation}
r_D=\sqrt{\frac{kT}{4\pi e^2 n_e}}\label{rD}
\end{equation}
For small values of ratio $\frac{r}{r_D}$, the potential can be
expanded into a series

\begin{equation}
\varphi(r)=\frac{e}{r}-\frac{e}{r_D}+...\label{rr}
\end{equation}
The following terms are converted into zero at $r\rightarrow 0$.
The first term of this series is the potential of the considered
particle. The second term is a potential induced by other
particles of plasma on the charge under consideration. And so the
correlation energy of plasma is \cite{1}

\begin{eqnarray}
\delta E_{corr}=-N_e\frac{\pi^{1/2}e^3
kT}{n_e}\biggl[Z^2\frac{n_n}{kT} +\biggl(\frac{\partial
n_e}{\partial \mu_e} \biggr)_{N,T}\biggr]^{3/2},\label{Fcor}
\end{eqnarray}
where $Z$ is the nuclear charge and $n_n=\frac{n_e}{Z}$ is the
density of nuclei.

Since the chemical potential of the Boltzmann ideal gas at high
temperature

\begin{equation}
\frac{d\mu_e}{dn_e}=\frac{kT}{n_e}
\end{equation}
and

\begin{equation}
\delta E_{corr}=-N_e \biggl(\frac{\pi
n_e}{kT}\biggr)^{1/2}(Z+1)^{3/2}e^3
\end{equation}

Since the space correlation an attraction between unlike charges
is prevalent over a repulsion of like charges, the plasma pressure
is below the pressure of the ideal gas at the same parameters. By
this reason this correction has negative sign.

\subsubsection{The density of hot dense plasma at the equilibrium state}

Finally,the full energy of plasma  in consideration of both main
corrections on inter-particle interaction given by

\begin{equation}
E=\frac{3}{2}kTN_e\biggl[1+\frac{\pi^{3/2}}{4}\biggl(\frac{a_0
e^2}{kT}\biggr)^{3/2}n_e - \frac{2\pi
^{1/2}}{3}\biggl(\frac{Z+1}{kT}\biggr)^{3/2}e^3 n_e^{1/2}\biggr]
\end{equation}

At a constant full number of particles in the system and at a
constant temperature, the equilibrium state exists at the minimum
of energy

\begin{equation}
\biggl(\frac{\partial E}{\partial n_e}\biggr)_{N,T}=0,\label{dedn}
\end{equation}
what allows one to obtain the steady-state value of density of hot
non-relativistic plasma

\begin{equation}
n_{\star}=\frac{16(Z+1)^{3}}{9 \pi^2 a_0^3}\simeq 2\cdot
10^{24}(Z+1)^3 ~ cm^{-3}, \label{neq}
\end{equation}
The Fermi-energy of electron gas of equilibrium plasma at this
density is

\begin{equation}
\varepsilon_F(n_{\star})= \biggl(\frac{16}{3}\biggr)^{2/3}\frac{m
e^4}{2\hbar^2}(Z+1)^2\approx 1.5~\frac{e^2}{a_0}(Z+1)^2\label{eF}
\end{equation}

\subsection{Equilibrium temperature of a hot non-relativistic
star} As the steady-state value of the density of hot
non-relativistic plasma is known, we can obtain a steady-state
value of the temperature of hot non-relativistic plasma.

According to the virial theorem \cite{1,BV+Lub}, the potential
energy $U$ of particles with Coulomb interaction is equal to their
double kinetic energy $T$ with opposite sign

\begin{equation}
U =-2T
\end{equation}
and their full energy is equal to  kinetic energy with opposite
sign. Neglecting small corrections at a high temperature, one can
write the full energy of hot dense plasma as

\begin{equation}
E_{plasma}= U + \frac{3}{2}kTN_e = - \frac{3}{2}kT N_e.
\end{equation}
As the plasma temperature is high enough, the pressure of the
black radiation cannot be neglected. The full energy of a star
depending on the plasma energy and the black radiation energy is

\begin{equation}
E_{total}=-\frac{3}{2}kT N_e + \frac{\pi^2}{15}
\biggl(\frac{kT}{\hbar c}\biggr)^3 V kT
\end{equation}
The equilibrium temperature of a body consisting of hot
non-relativistic plasma is determined by the energy minimum
condition

\begin{equation}
\biggl(\frac{\partial E_{total}}{\partial T}\biggr)_{N,V} =0.
\end{equation}
It gives the following value of the equilibrium temperature

\begin{equation}
T_{\star}=\biggl(\frac{10}{\pi^4}\biggr)^{1/3}(Z+1)\frac{\hbar
c}{ka_0}\approx 2  \cdot 10^7(Z+1)~K.\label{4a}
\end{equation}

All substances usually have a positive thermal capacity. Therefore
the minimal energy for such substances exists at $T=0$. The
existence, in our case, of the energy minimum at the finite
temperature $T_{\star}\neq 0$ is not confusing. Each small part of
a star has a positive thermal capacity, but a gravitational
interaction of these parts between themselves results in a
situation where the thermal capacity of a star as the whole
becomes negative at some temperature and a star energy decreases
with an increased temperature. As a result there are two branches
of the temperature dependence of a star energy - with the negative
capacity at low temperatures and with a positive capacity at high
temperatures. Between them, at some finite temperature $T_{\star}$
there is the minimum of the energy.

The steady-state values of the density and the temperature for hot
non-relativistic plasma have been considered above. One can see,
that the criterion of hot plasma
\begin{equation}
kT \gg \varepsilon_F
\end{equation}
is satisfied according of Eq.{\ref{eF}} if the nuclear charge is
not too large

\begin{equation}
\frac{kT_{\star}}{\varepsilon_F(n_{\star})}=\biggl(\frac{45}{16\pi^4}\biggr)^{1/3}~\frac{\hbar
c}{(Z+1)e^2} \approx \frac{0.3}{Z+1}~\alpha^{-1}\approx
\frac{40}{Z+1},\label{ktef}
\end{equation}
where $\alpha=e^2/\hbar c=1/137$ is the fine structure constant.

\section{The equilibrium of a dense electron-nuclear
plasma}

According to the definition (Eq.({\ref{muB}})) plasma chemical
potential in equilibrium state at constant temperature and density
must also be constant:

\begin{equation}
\mu(n_{\star},T_{\star})=const
\end{equation}
or
\begin{equation}
\nabla \mu = 0\label{m0}
\end{equation}

How  can one obtain an agreement between this condition and the
equation of equilibrium in gravitation field (Eq.({\ref{equ1}}))?

The condition of the equilibrium of plasma in gravity field can be
obtained if the role of non-gravitational fields, for example, the
electric field, is taken into account. More precisely, the
equilibrium equation (Eq.({\ref{equ1}})) must include all fields,
which can have an impact on particles, for example, electric field
for system of charged particles:
\begin{equation}
\mu+\sum_i q_i\phi_i=const
\end{equation}
(where $q_i$ and $\phi_i$ are charge and potential of any nature
(gravitational, electric)).

In a spherical symmetric case at $\mu=const$ it reduces to

\begin{equation}
G\gamma m'=\rho q
\end{equation}
where $\rho=q\cdot n_e$ is the electric charge density and
\begin{equation}
q=G^{1/2}m'\label{q}
\end{equation}
is the charge induced in plasma cell by gravity (related to one
electron). One should not think that the  gravity field really
induces in plasma some additional charge. We can rather  speak
about electric polarization of plasma that can be described as
some redistribution of internal charges in the plasma body.

Essentially it stays electrically neutral as a whole, because the
positive charge with volume density

\begin{equation}
\rho=G^{1/2}{\gamma}
\end{equation}
is concentrated inside the charged plasma core and the
corresponding negative electric charge exists on its surface.
Since
\begin{equation}
4\pi\rho=div \mathbf{E}\label{divE}
\end{equation}
and
\begin{equation}
-4\pi G \gamma=div \mathbf{g}\label{divg}
\end{equation}
the equilibrium equation can be rewritten as

\begin{equation}
\gamma \mathbf{g}+\rho \mathbf{E}= \mathrm{0}\label{f1}
\end{equation}
where

\begin{equation}
{\mathbf{E}}=\frac{{\mathbf{g}}}{G^{1/2}}.
\end{equation}

It must be noted, the using of Thomas-Fermi approximation gives
possibility to consider the balance in plasma cells in more detail
\cite{BV}.

\subsubsection{The equilibrium density of another kind of dense plasmas}

The above consideration of equilibrium of the hot non-relativistic
plasma in a self-gravity is characterized by its obviousness, but
the similar equilibrium is not the characteristic property for
this kind of plasma only.

The direct consideration of the plasma equilibrium in Fermi-Thomas
approximation shows that the application of a gravity field to
plasma induces its electric polarization
(\cite{BV},\cite{BV-book}), so the equilibrium equation in the
form Eq.({\ref{f1}}) is applicable to all kind of dense plasmas -
relativistic or non-relativistic and simultaneously degenerate or
non-degenerate ones.

It is essential that one can find  the constancy of density and
chemical potential of another kind of plasmas in an equilibrium
condition.

For a cold non-relativistic plasma the kinetic energy of electron
is

\begin{equation}
E_k=\frac{3}{5}E_F=\frac{3}{10} (3\pi^2)^{2/3}a_0 e^2 n_e^{2/3}.
\end{equation}

Its potential energy is

\begin{equation}
E_p\approx - e^2 n_e^{1/3}.
\end{equation}
According to the virial theorem $E_k\approx -E_p$. Thus the
equilibrium electron density is

\begin{equation}
n_e \approx a_0^{-3}.
\end{equation}
and it does not depend on temperature.

The relativistic plasma exists at a huge pressure which induces
the neutronization of substance. For this process the density

\begin{equation}
n_e=\frac{\triangle^{3}}{3\pi^2 (c\hbar)^3} \label{neu}
\end{equation}
is characteristic \cite{1}. (Where $\triangle$ is the difference
of nuclear bonding energy of neighbouring interacting nuclei.)
Since the difference

\begin{equation}
\triangle \approx m_e c^2\label{ne1}
\end{equation}
the equilibrium of relativistic plasma density (at condition of a
homogeneous mixing of reacting substance)

\begin{equation}
n_e\approx \frac{1}{3\pi^2}(\alpha a_0)^{-3} \approx 10^{30},
cm^{-3}\label{ne2}
\end{equation}
where $\alpha=e^2/\hbar c$ is the fine structure constant,
$a_0=\hbar^2/m_ee^2$ is the Bohr radius.

A similar consideration can be extended on the neutron matter if
 it is considered as electron-proton plasma in the neutron
environment.

\section{The giro-magnetic ratio of stars}

Gravitation produces a redistribution of free charges in
plasma inside a star. Essentially, a star as a whole conserves its
electric neutrality. However, as the star rotates about its axis,
positive volume charges are moving on smaller radial distances
than the surface negative charge. It induces a magnetic field
which can be measured.

The magnetic moment of the surface spherical layer, which carries
the charge $Q$, is

\begin{equation}
\mu=\frac{Q\Omega R^2}{3c},
\end{equation}
where $\Omega$ is rotational velocity, $Q=\frac{4\pi}{3}\rho R^3$.

The magnetic moment induced by a volume charge is small because it
is concentrated in the small central core for the most part.

On the other hand, the angular momentum of a star is approximately

\begin{equation}
L\approx\frac{2}{5}M\Omega R^2
\end{equation}
and the giro-magnetic ratio of a star is expressed through the
world constants only:

\begin{equation}
\vartheta\approx\frac{\mu}{L}\approx\frac{\sqrt{G}}{c}\label{3a}
\end{equation}
It can be verified by the measurement data.

\begin{figure}
\includegraphics[5cm,8cm][9cm,18cm]{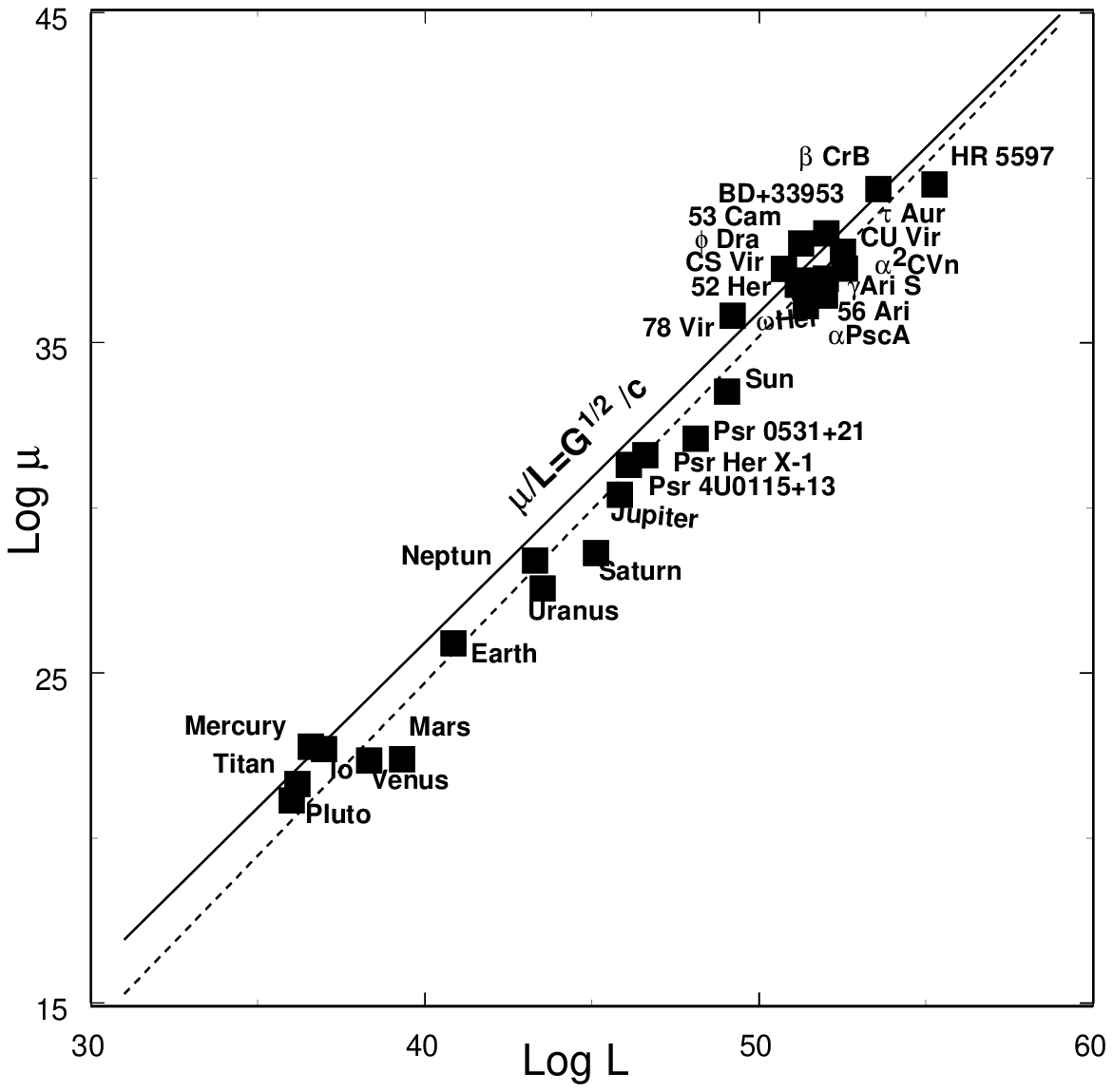}
\vspace{6cm} \caption { The observed values of the magnetic
moments of celestial bodies vs. their angular momenta. On the
ordinate, the logarithm of the magnetic moment over
$Gs\cdot{cm^3}$ is plotted; on the abscissa the logarithm of the
angular momentum over $erg\cdot{s}$ is shown. The solid line
illustrates  Eq.({\ref{3a}}). The dash-dotted line is the fitting
of the observed values.} \label{mL}
\end{figure}

The values of giro-magnetic ratio for all celestial bodies (for
which they are known today) are shown in Fig.{\ref{mL}}.

The data for planets are taken from \cite{Sirag}, the data for
stars are taken from \cite{Borra}, and  those for pulsars - from
\cite{Beskin}. Therefore, for all celestial bodies - for planets
and their satellites, for $Ap$-stars and several pulsars, which
angular momenta themselves change within the limits of more than
20 orders - calculated values of the gyromagnetic ratio
Eq.({\ref{3a}}) with a logarithmic accuracy quite satisfactorily
agrees with measurements.

\section{The stellar mass distribution}
\subsection{The mass of star consisting of hot non-relativistic dense plasma}

Inside the stellar core consisting of hot dense plasma the
gravitational force is counterbalanced by the electric force
(Eq.({\ref{f1}})) and there is no gradient of pressure. The
absence of a pressure gradient inside the hot star core does not
mean the absence of pressure. It is not  difficult to show that if
the gravity force inside a star is compensated by the electric
force, the negative energy of gravitational field is canceled by
the energy of the electric field. The non-compensated part of the
total energy is the energy of gravitational field outside a star.
This field has the energy

\begin{equation}
E_G=-\frac{GM_{\star}^2}{2R_{\star}}
\end{equation}
where $M_{\star}$ and $R_{\star}$ are the mass and the radius of
the hot plasma core. This external gravitational field endeavoures
to compress a star.

The virial theorem connects the gravitational energy of a star and
pressure inside it \cite{1}
\begin{equation}
E_G=-3\int PdV,
\end{equation}
or in our case (P=const)
\begin{equation}
E_G=-3PV_{\star},\label{tv}
\end{equation}
where $V_{\star}=\frac{4\pi}{3}R_{\star}^3$ is the volume of the
star core. Taking into account of the black radiation pressure we
obtain
\begin{equation}
\frac{GM_{\star}^2}{6RV}=kT_{\star}n_{\star} + \frac{\pi^2}{45}
\frac{(kT_{\star})^4}{(\hbar c)^3}
\end{equation}
Finally, for  the mass of the star core consisting from hot dense
plasma we have
\begin{equation}
M_{\star}=\frac{5^{1/2}3^3}{2\pi^{3/2}} \biggl(\frac{\hbar c}
{Gm_p^2}\biggr)^{3/2} \biggl(\frac{Z}{A}\biggr)^2 m_p\nonumber\\
\approx 5.42 M_{Ch} \biggl(\frac{Z}{A}\biggr)^2, \label{4b}
\end{equation}
where $M_{Ch}=\biggl(\frac{\hbar c}{G
m_p^2}\biggr)^{3/2}m_p=3.42\cdot 10^{33} g$ is the Chandrasechar
mass. It is important to underline that the obtained stellar mass
estimation (Eq.({\ref{4b}})) is depending  only on one variable
parameter $A/Z$.

One can note that the equilibrium radius of the star core

\begin{eqnarray}
R_{\star}= \frac{(3/2)^3}{2}
\biggl(\frac{10}{\pi}\biggr)^{1/6}\biggl(\frac{\hbar c}{G
m_p^2}\biggr)^{1/2} \frac{a_0}{(Z+1){A/Z}}.\label{s10}
\end{eqnarray}

also depends on variables $A$ and $Z$ only.

\subsection{The mass of a star consisting of cold relativistic plasma}

With then increase of the density, the plasma can turn into a
relativistic state. It occurs when Fermi momentum of electrons
satisfies
\begin{equation}
p_F=(3\pi^2)^{1/3}n_e^{1/3} \hbar>m_e c
\end{equation}


This value of momentum corresponds to the steady-state density of
a substance under neutronization (Eq.({\ref{ne2}})) at
$n_{\ast}\approx 10^{30} ~cm^{-3}$. At  temperature

\begin{equation}
T\ll\frac {mc^2} {k} \approx 10^{10}~K.
\end{equation}
it can be considered as cold.

As the pressure of the relativistic electron gas is
\begin {equation}
P_R=\frac{(3\pi^2)^{1/3}}{4} n^{4/3}\hbar c,
\end {equation}
according to Eq.({\ref{tv}}), the pressure balance obtains the
form:

\begin {equation}
\frac{GM_{\ast}^2}{6R_{\ast}V_{\ast}}=\frac{(3\pi^2)^{1/3}}{4}
n^{4/3}\hbar c,
\end {equation}

Therefore, the relativistic degenerate star in equilibrium state
must have the steady value of mass

\begin {equation}
M_{\ast}= {1.5}^{5/2}\pi^{1/2}\biggl(\frac{\hbar c}{Gm_p^2}
\biggr)^{3/2}\frac{m_p}{(A/Z)^2}\approx
\frac{4.88M_{Ch}}{(A/Z)^2}\label{5b}
\end {equation}
at the radius corresponding to Eq.({\ref{ne2}})

\begin {equation}
R_{\ast}\approx\biggl(\frac{\hbar c} {G m_p^2}\biggr)^{1/2}
\frac{\alpha a_0}{A/Z}\approx \frac{10^{-2}R_{\odot}}{(A/Z)}
\end {equation}

The objects which have such masses and density are best suited to
the dwarfs.

\subsection{The comparison of calculated star masses with observations}

\begin{figure}
\begin{center}
\includegraphics[10cm,3cm][9cm,12cm]{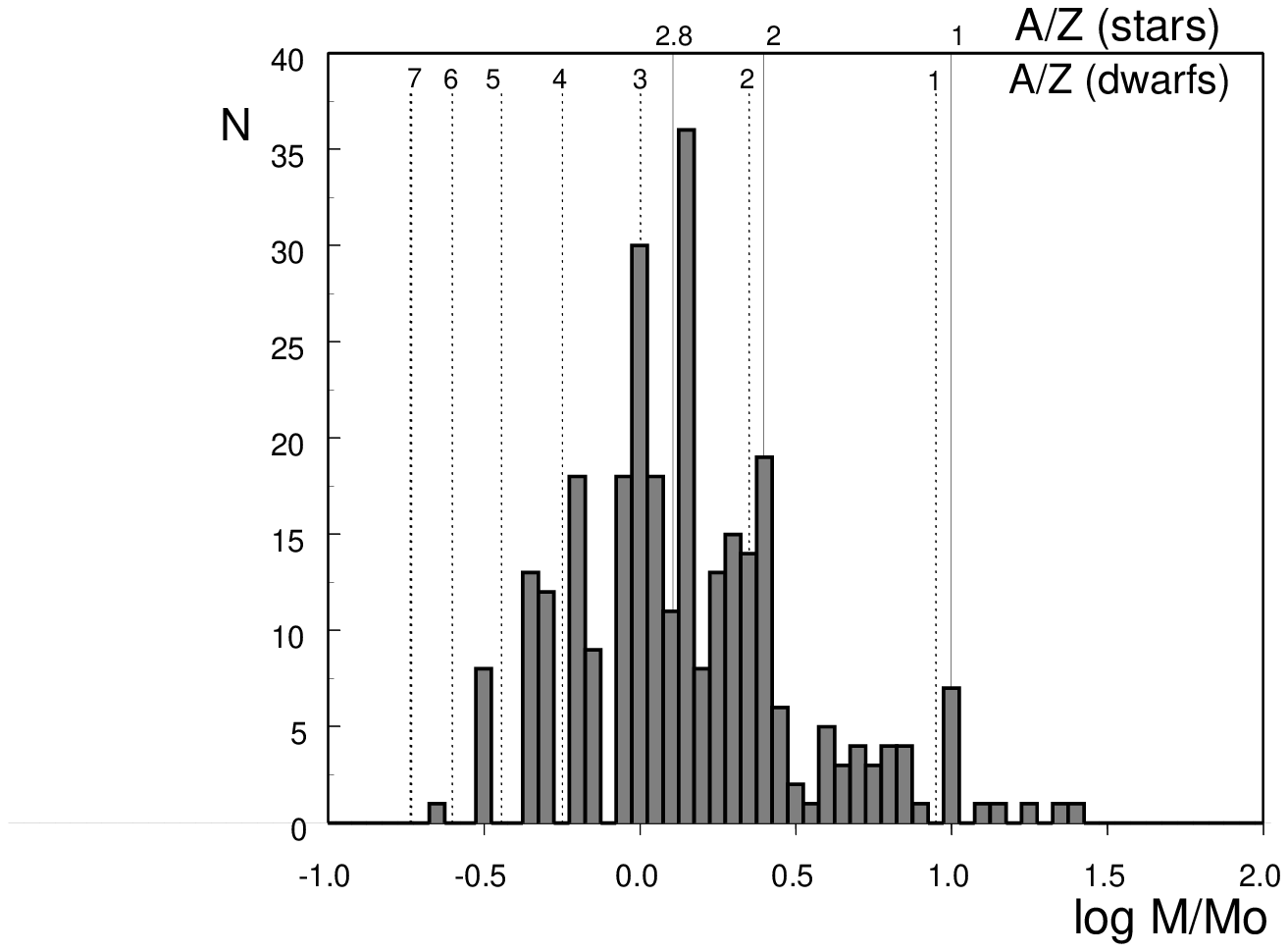}
\caption{ The mass distribution of  binary stars \cite{Heintz}. On
abscissa, the logarithm of the star mass over the Sun mass is
shown. Solid lines mark masses  which agree with selected values
of A/Z from Eq.({\ref{4b}}) for stars. The dotted lines mark
masses which agree with selected values of A/Z from
Eq.({\ref{5b}}) for dwarfs.} \label{stars}
\end{center}
\end{figure}

The comparison of calculated results with the data of measurements
is shown in Fig.{\ref{stars}}. There is a large quantity of star
mass measurements but only those, which was obtained from the
measurement of binary star parameters, have a sufficient accuracy
only. The mass distribution of visual and eclipsing binary stars
\cite{Heintz} is shown in Fig.{\ref{stars}}. On abscissa, the
logarithm of the star mass over the Sun mass is plotted. Solid
lines mark masses which agree with selected values of A/Z for
stars from Eq.({\ref{4b}}). The dotted lines mark A/Z for dwarfs
from Eq.({\ref{5b}}).

According to the existing knowledge, the hydrogen inside dwarfs is
fully burnt out. In full agreement with it in Fig.{\ref{stars}}
there are hydrogen stars and there are not dwarfs with $A/Z=1$,
whereas there are both - helium-deuterium stars and dwarfs - with
$A/Z=2$. Attention should be attracted to the fact that there is
the peak in the star distribution consisting of heavy nuclei with
$A/Z=2.8$. The nuclei with $A/Z > 2.8$ are absent in the
terrestrial condition. This ratio of $A/Z$ is the limit value for
the stable nuclei at a relatively small pressure. One can expect
that at high pressure which can induce the neutronization process,
the neutron-excess nuclei with large $A/Z$ ratio obtain stability.
Attention is attracted to the peak for dwarfs with $A/Z=3$ where
the Sun is placed. It seems that stars composed of nuclei with
$A/Z$ ratio up to 7 are presented in this spectrum however this
question is complicated and demands a special and more attentive
consideration.

\subsection{Masses of star cores, composed by another kinds of plasma.}

Let us take a quick look on star cores composed by another kinds
of plasma. As before we will proceed from the virial theorem ratio
({\ref{tv}}).

\subsubsection{Celestial bodies composed by cold non-relativistic plasma}

The cold plasma exists at temperatures
\begin{equation}
T<<T_F \label{25010}
\end{equation}
and has a pressure
\begin{equation}
P=\frac{(3\pi^2)^{2/3}}{5} \frac{\hbar^2}{m}
\biggl(\frac{\gamma}{m_p A/Z}\biggr)^{5/3}.\label{pl10}
\end{equation}

It gives a possibility to represent the equilibrium equation
({\ref{tv}}) as
\begin{equation}
\frac{GM^2}{2R}=3V\frac{(3\pi^2)^{2/3}}{5} \frac{\hbar^2}{m}
\biggl(\frac{\gamma}{m_p A/Z}\biggr)^{5/3}\label{pl11}
\end{equation}
and to obtain the expression for a mass of the body

\begin{equation}
M=M_{Ch} \biggl(\frac{\hbar}{mc}\biggr)^{3/2}
\biggl(\frac{\gamma}{m_p}\biggr)^{1/2}
\frac{6^{3/2}9\pi}{4(A/Z)^{5/2}}\label{pl12}
\end{equation}
As  the  degenerate non-relativistic plasma has the density
$\gamma\approx 1~g/cm^3$, we obtain

\begin{equation}
M\approx 1.26\cdot 10^{-3} \frac{M_{Ch}}{(A/Z)^{5/2}}
\approx\frac{4.3\cdot 10^{30}}{(A/Z)^{5/2}} ~ g\label{pla}
\end{equation}

At this density, the degeneration temperature $T_F\approx 10^5K$.
Thus the temperature of the object under consideration should not
exceed several thousands degrees. Among celestial bodies, only
planets possess these properties.

\begin{figure}
\begin{center}
\includegraphics[3cm,6cm][9cm,12cm]{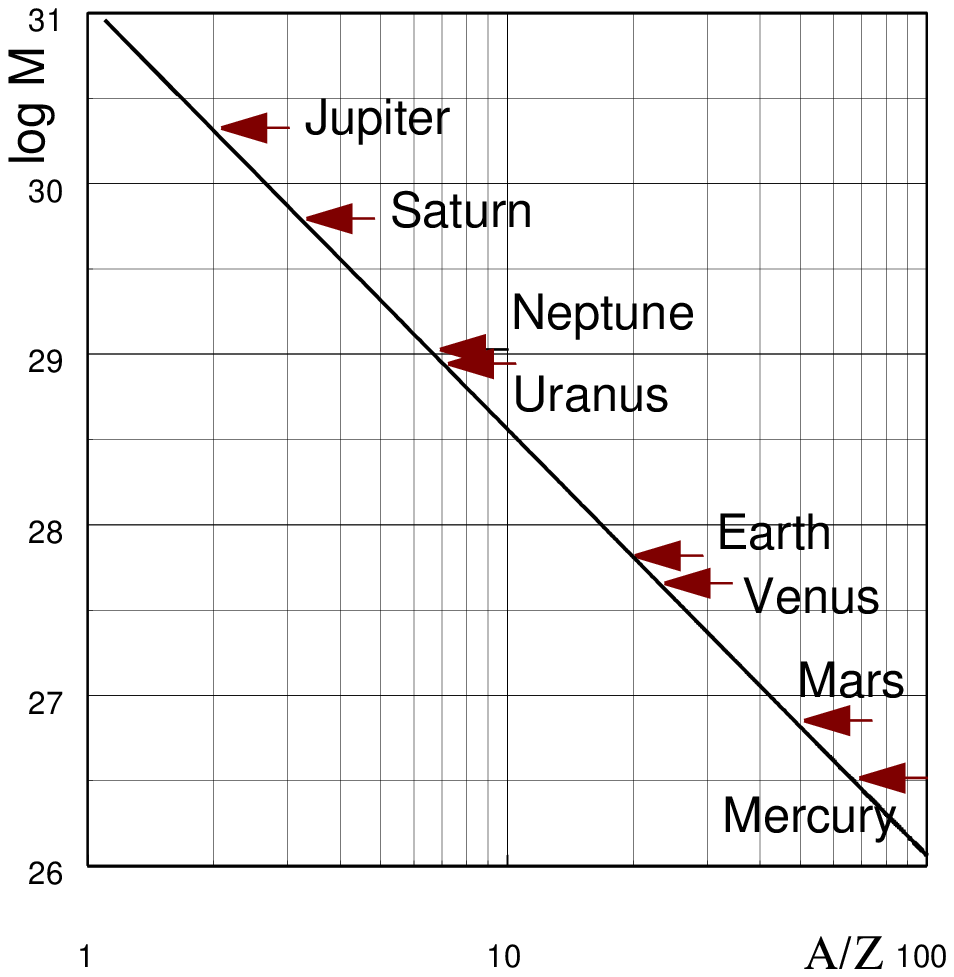}
\vspace{5cm} \caption{The dependence of the core mass of planets
over ratio $A/Z$ (Eq.({\ref{pla}})) at $\gamma =1~g/cm^{3}$. On
the ordinate, the logarithm of mass (over $1~g$) is plotted.}
\label{planet}
\end{center}
\end{figure}

The comparison of the obtained estimation ({\ref{pla}}) and the
measured data for Solar system planets is shown in
Fig.(\ref{planet}). It is seen that the obtained estimation is in
agreement with the  value of masses of large planets at
$A/Z\approx 2$. It is important to note that according to
Eg.(\ref{pla}), planets wiht masses more than $10^{31}g$ must not
exist. In reality only the Jupiter has the mass of this level.

\subsubsection{Do stars consisting of hot ultra-relativistic plasma
represent quasars?}

A star may be considered as a hot one if the temperature of
radiation inside it is much higher than the degeneration
temperature of its electron gas:
\begin{equation}
\frac{T_R}{T_F}\gg 1
\end{equation}
For non-relativistic star, this ratio is  approximately equal to
$1/3\alpha$ (Eq.({\ref{ktef}})). At this condition, the  pressure
of degenerate electron gas can be neglected
\begin{figure}
\begin{center}
\includegraphics {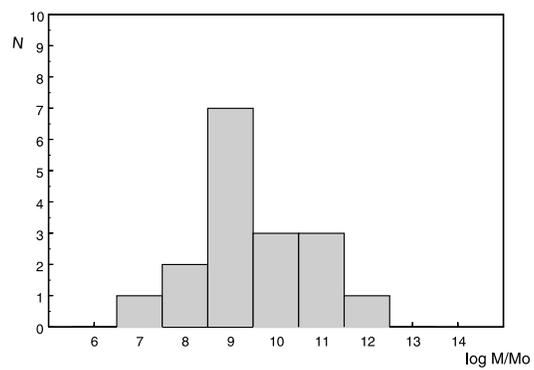}
\caption{The mass distribution of galaxies \cite{Allen}. In
abscissa, the logarithm of the galaxy mass over the Sun mass is
shown.} \label{galaxy}
\end{center}
\end{figure}

and  the equilibrium equation takes the form

\begin{equation}
\frac{GM^2}{6RV}=\frac{\pi^2}{45}\frac{(kT_R)^4}{(\hbar
c)^3}\approx \biggl(\frac{T_R}{T_F}\biggr)^3 kT_Rn\label{qu1}
\end{equation}

It gives a possibility to estimate the mass of a hot
ultra-relativistic star

\begin{equation}
M\approx \Biggl(\frac{T_R}{T_F}\Biggr)^6 M_{Ch} \label{qu2}
\end{equation}

According to the existing knowledge, among compact celestial
objects only quasars have masses of this level.

Apparently it is an agreed-upon opinion that quasars represent a
relatively short stage of the evolution of galaxies. If to adhere
to this hypothesis and because of the lack of information about
quasar mass distribution, we can use the distribution of masses of
galaxies to check our estimation (Fig.{\ref{galaxy}).

It can be seen that obtained quasar equilibrium conditions
(Eq.({\ref{qu1}}),Eq.({\ref{qu2}})) is in agreement with the
observation data at
\begin{equation}
10<\frac{T_R}{T_F}<100\label{nr2}
\end{equation}
and, as for non-relativistic hot stars, some maximum exists at the
equilibrium condition of (Eq.({\ref{ktef}})):
\begin{equation}
\frac{T_R}{T_F}\approx \frac{1}{3\alpha}\label{r22}
\end{equation}

Thus, the conclusion that quasars consist of a hot relativistic
plasma is not contradict with the observations.

\subsubsection{The mass of a star consisting of neutron matter}
Dwarfs and quasars may be considered as stars where the process of
neutronisation is just beginning. Finally, at nuclear density,
plasma turns into neutron matter.

It is agreed that a pulsar is a star consisting of neutron matter
with some impurity of other particles - electrons and protons.
Evidently, at nuclear density neutrons and protons are
indistinguishable inside pulsars as well as inside a huge nucleus.
The neutron matter can be considered as a kind of  plasma where
electrons and protons exist in a neutron environment. Under this
condition a very small impurity of electrons and protons (about a
level of $10^{-18}$) is enough to induce a sufficient electric
polarization and to balance the action of gravity.

It is known from nuclear physics, that a nuclear matter is
incompressible one. By this reason one can expect that its mass
density inside neutron star, similarly as for atomic nuclei,
approximately equals $3\cdot10^{14} g/cm^3$ at partial density
$1.8\cdot 10^{38} cm^{-3}$. At this density it may be considered
as cold at $T\ll 10^{12}K$. It is important  to note that the
neutron gas is not ultra-relativistic at the nuclear density. It
is a relativistic gas only when the Fermi momentum of neutrons is
\begin{equation}
\frac{p_F}{m_p c}\approx 0.36 .
\end{equation}
The pressure of this neutron gas is described by the complicated
equation in general case \cite{1}:
\begin{equation}
P=\frac{m^4 c^5}{32\pi^2 \hbar^3} \biggl[\frac{1}{3} sh
 \xi-\frac{8}{3}sh \frac{\xi}{2}+\xi\biggr],\label{13270}
\end{equation}
where $\xi=4 Arsh\frac{p_F}{mc}$.

The equilibrium equation
\begin{equation}
\frac{GM_{pulsar}^2}{6RV}=\frac{m_p^4 c^5}{32 \pi^2
\hbar^3}\biggl(\frac{1}{3} sh\xi
-\frac{8}{3}sh\frac{\xi}{2}+\xi\biggr)
\end{equation}
gives the equation for the neutron star mass
\begin{equation}
M_{pulsar}= \frac{3^4 \sqrt{\pi}}{2^7} M_{Ch} F , \label{6a}
\end{equation}
where
\begin{equation}
F=\biggr[\frac{\frac{1}{3}sh\xi-\frac{8}{3}sh\frac{\xi}{2}+\xi}
{\bigl(sh\frac{\xi}{4}\bigr)^4}\biggr]^{3/2} . \label{6b}
\end{equation}
At $\frac{p_F}{m_p c}\approx 0.36$ we have $F\approx 0.64$ and
\begin{equation}
M_{pulsar}= 1.32 M_{\odot}, \label{6c}
\end{equation}
where $M_{\odot}$ is the Sun mass. This estimation is in a good
agreement with the pulsar mass measuring data \cite{Thorsett},
which is shown in Fig.({\ref{pulsar}}). In the upper scale the
density of neutron matter according to
Eqs.({\ref{6a}})-({\ref{6b}}) is plotted. One can see that the
result of the  pulsar masses measurement is in full agreement
with the incompressible nuclear matter hypothesis.

\clearpage

\begin{figure}
\begin{center}
\includegraphics[4cm,1cm][10cm,11cm]{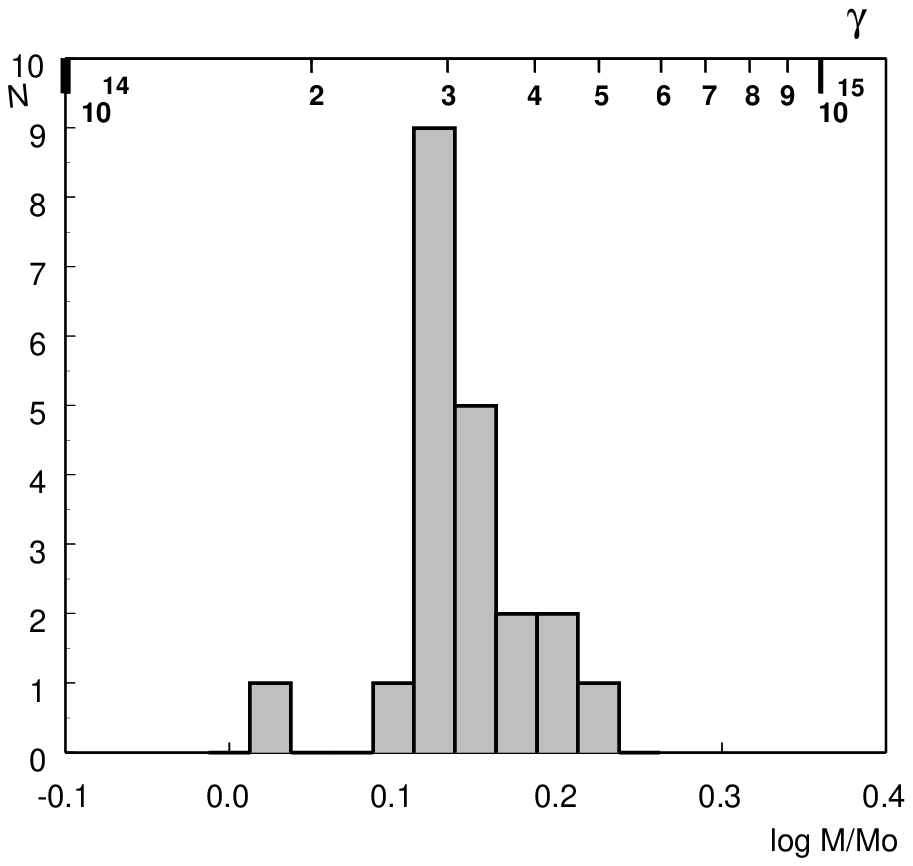}
\caption { The mass distribution of pulsars \cite{Thorsett}. In
lower abscissa, the logarithm of the pulsar mass over the Sun mass
is shown. In upper abscissa, the density of the substance in
$g\cdot cm^{-3}$ according to Eqs.({\ref{6a}})-({\ref{6b}}) is
plotted.} \label{pulsar}
\end{center}
\end{figure}

\clearpage

\section{The classification of stellar objects}

The commonly accepted classification of stars is based on their
external indicators: surface temperature,  luminosities,
characteristic properties of radiation. That is why the
classification of stars according to equation of states of their
substance may represent a more physical, stringent and consistent
method. First of all, this makes evident in the determination of
the number of classes into which all celestial objects of the
Universe may be divided.

The matter may exist in seven states.

Atomic substances may have

1. A condensed state (solid or liquid) at low temperatures.

2. A gas state at a high temperature.

The electron-nuclear plasma may have four states. They are

3. The non-relativistic and simultaneously degenerate (cold)
plasma.

4. The non-relativistic and non-degenerate (hot) plasma.

5. The relativistic and degenerate (cold) plasma.

6. The relativistic hot plasma, consisting of the relativistic
electron gas and the radiation which temperature is larger than
the degeneration temperature of the electron gas.

And furthermore

7. The neutron matter with the nuclear density can have degenerate
state.

Nowadays, assumptions of the existence of the substance in a state
other than indicated above seem to be unjustified. Consequently,
the classification of all celestial bodies in accordance with
possible states of substance should be performed dividing these
objects into seven classes.

1. If  the mass of a celestial body is  relatively small, the
pressure at its central region is also small and plasma is absent
there. In this case, the whole body consists of an atomic
substance. Small celestial bodies like asteroids and satellites of
the planets are related to this class. At higher mass, a
non-relativistic degenerate ion-electron plasma core exists in the
central region of a body. This core is covered by a mantle of
atomic substance in condensed form. This picture relates to
relatively small planets (like Earth)  and some of their
satellites.

2. If the temperature is sufficient for evaporation but lower than
the ionization temperature,  a celestial body is consist of a gas.
Because this body can be formed as a result of a the cooling of a
hot star, it may have a mass which is typical for a star with the
gas density and a relatively small temperature. The very large
dimensions and the small density give a possibility to classify
giants into this class.

3. Large planets can be considered as objects consisting of
non-relativistic and degenerate plasma. Atomic mantles play an
unimportant role in their forming. The equilibrium equation leads
to the limitation of their masses ({\ref{pla}}).

4. Stars consisting of hot non-degenerate non-relativistic plasma
must have the internal temperature  $\approx \hbar c / a_0 k
\approx 10^7K$ and the masses $\approx 10^{33}$g, depending on
their nuclear composition.

5. Dwarfs, consisting of relativistic degenerate plasma and having
high density cores with substance under neutralization, must have
small radii. Their states (at temperatures $T<mc^2/k\approx
10^{10}K$) are not depending on their temperatures and can be
considered as steady.

6. Quasars are consist of the relativistic plasma and the high
density radiation. They can exist in steady state at masses
$\approx 10^6 \div 10^{12} M_{Ch}$ and temperatures
$T>mc^2/k\approx 10^{10}K$.

7. Pulsars consisting of neutron matter with nuclear density
$3\cdot 10^{14} g/cm^3$ at $\frac{T_R}{T_F}<1$ can be considered
as bodies in steady state too.

One can expect that some transitional and intermediate states can
exist. They fall out of this static classification, because it
classifies objects in steady state only. But a static character of
the classification does not prevent to review  some dynamics of
stellar matter.

It seems that there is no thermodynamical prohibition to suppose
the existence  a stellar object consisting of neutron matter or
relativistic plasma with radiation at a temperature $T>>10^{12}~K$
at some starting phase. The equilibrium mass of this strange
object must be on the level $10^{53}$g, i.e. all mass of the
reviewing Universe can be concentrated in this compact body in the
starting phase. The following cooling must disturb its equilibrium
and it can decay in approximately $10^{10}$ quasars which
steady-state temperatures are approximately two orders below. A
cooling of quasars leads to the loss of their equilibrium and to
their decay on galaxies of hot stars with masses about $M_{Ch}$
and temperatures about $10^7$K. The following cooling of stars may
lead to the birth of pulsars or dwarfs, or to the scattering of
the hot star substance in small cool celestial bodies - planets,
asteroids or in gas clouds, which are stable on this stage of
cooling and  expansion of Universe.

\section{Conclusion.}\vspace{12pt}
The results obtained above sharply distinguish from the point of
view which is commonly accepted in the astrophysical society. It
is important to note that no disputable speculative assumptions
were made above. All these results were obtained on the standard
physical base by standard formal methods. The novelty of the
developed approach to fundamental astrophysical problems is based
on a rejection of the usually accepted stating that an increase of
the temperature and the density depthward of a celestial body is a
requirement of the equilibrium of a substance in the self-gravity
field. This requirement is really applicable to the equilibrium of
atomic substances and does not applicable to plasma. It is shown
above that there is the  minimum for electron-nuclear plasma
energy at a density and a temperature which determines the
equilibrium state of plasma in the self-gravitation at the zero
gradient of the general parameters of plasma.

This effect provides the simple mechanism of the generation of the
magnetic field by celestial bodies. It can be noted that all
previous models tried to solve the other basic problem: they tried
to calculate the magnetic field of a celestial body. Now space
flights and the development of astronomy discovered a remarkable
and previously unknown fact: the magnetic moments of all celestial
bodies are proportional to their angular momenta and the
proportionality coefficient is determined by the ratio of world
constants only (\cite{Blackett},\cite{Sirag}). Nowadays the
explanation of this phenomenon is really the basic problem of
planetary and stars magnetism. The  theory developed in this paper
gives a simple and standard solution to this problem.

Our approach gives a possibility to predict important properties
of stars in their steady state. Starting from the equilibrium
conditions it allows to calculate masses of different types of
stars. Thus the masses of stars composed by non-relativistic
non-degenerate plasma and dwarfs composed by relativistic
degenerate plasma can be expressed by the ratio of world constants
and one variable parameter $(A/Z)$ only, and this statement is  in
a rather good agreement with the observation data. Just as the
predicted value of mass of pulsars is in full agreement with
observations at the assumption of incompressibility of nuclear
matter.

Some considered questions, especially the electric polarization of
plasma in a self-gravitational field, are analyzed more
systematically in \cite{BV},\cite{BV-book}.

\end{document}